\documentstyle[twoside,12pt,epsf]{article}
\input epsf
\epsfverbosetrue

\def\halfspace {\smallskipamount=6pt plus2pt minus2pt
                  \medskipamount=12pt plus4pt minus4pt
                  \bigskipamount=24pt plus8pt minus8pt
                  \normalbaselineskip=16pt plus0pt minus0pt
                  \normallineskip=2pt
                  \normallineskiplimit=0pt
                  \jot=6pt
                  {\def\smallskip {\vskip\smallskipamount}}
                  {\def\medskip   {\vskip\medskipamount}}
                  {\def\bigskip   {\vskip\bigskipamount}}
                  {\setbox\strutbox=\hbox{\vrule
                    height17.0pt depth7.0pt width 0pt}}
                  \parskip 12.0pt
                  \normalbaselines}

\def\pprintspace {\smallskipamount=4pt plus1pt minus1pt
                  \medskipamount=9pt plus2pt minus2pt
                  \bigskipamount=16pt plus4pt minus4pt
                  \normalbaselineskip=14pt plus0pt minus0pt
                  \normallineskip=1pt
                  \normallineskiplimit=0pt
                  \jot=4pt
                  {\def\smallskip {\vskip\smallskipamount}}
                  {\def\medskip   {\vskip\medskipamount}}
                  {\def\bigskip   {\vskip\bigskipamount}}
                  {\setbox\strutbox=\hbox{\vrule
                   height9.5pt depth4.5pt width 0pt}}
                  \parskip 0pt
                  \normalbaselines}

\def\refitem{\par\noindent\hangindent 20pt}
\def\wisk#1{\ifmmode{#1}\else{$#1$}\fi}
\def\lt     {\wisk{<}}
\def\gt     {\wisk{>}}
\def\amin   {\wisk{^\prime\ }}
\def\deg    {\wisk{^\circ}}
\def\ddeg   {\wisk{{\rlap.}^\circ}}
\def\muK{\wisk{{\rm \mu K}}}

\begin{document}
\pagestyle{plain}
\pprintspace

% --------------------- Title --------------------- 
\large
\begin{center}
High-Latitude Galactic Emission \\
in the {\it COBE} DMR Two-Year Sky Maps
\end{center}

% --------------------- Author list --------------------- 
\medskip
\normalsize
\pprintspace
\noindent
\begin{center}
A.~Kogut\footnotemark[1]$^{,2}$,
A.J. Banday$^1$,
C.L. Bennett$^3$,
K.M. G\'{o}rski$^{4,5}$, \\
G. Hinshaw$^1$, 
and
W.T. Reach$^4$
\end{center}
\footnotetext[1]{
~Hughes STX Corporation, Laboratory for Astronomy and Solar Physics, 
Code 685, NASA/GSFC, Greenbelt MD 20771. \newline
\indent~$^2$ E-mail: kogut@stars.gsfc.nasa.gov. \newline
\indent~$^3$ Laboratory for Astronomy and Solar Physics, 
NASA Goddard Space Flight Center, Code 685, Greenbelt MD 20771. \newline
\indent~$^4$ Universities Space Research Association, 
Laboratory for Astronomy and Solar Physics, 
Code 685, NASA/GSFC, Greenbelt MD 20771. \newline
\indent~$^5$ On leave from Warsaw University Observatory,
Aleje Ujazdowskie 4, 00-478 Warszawa, Poland
\newline
}

\medskip
\normalsize
\pprintspace
\begin{center}
{\it COBE} Preprint 95-09 \\
Accepted for publication in {\it The Astrophysical Journal} \\
\end{center}

% -------------------- Abstract -------------------- 
% Keep abstract on same page as title
\medskip
\begin{center}
\large
ABSTRACT
\end{center}

\normalsize
\noindent
We cross-correlate the {\it COBE}\footnotemark[6]
\footnotetext[6]{
~The National Aeronautics and Space Administration/Goddard Space Flight Center
(NASA/GSFC) is responsible for the design, development, and operation of the
Cosmic Background Explorer ({\it COBE}).
Scientific guidance is provided by the {\it COBE} Science Working Group.
GSFC is also responsible for the analysis software
and for the production of the mission data sets.
}
DMR 2-year sky maps with spatial templates 
from long-wavelength radio surveys
and the far-infrared 
{\it COBE} DIRBE maps.
We place an upper limit on the spectral index of synchrotron radiation
$\beta_{\rm synch} < -2.9$
between 408 MHz and 31.5 GHz.
We obtain a statistically significant cross-correlation with the DIRBE maps
whose dependence on the DMR frequencies 
indicates a superposition of dust and free-free emission.
The high-latitude dust emission ($|b| > 30\deg$)
is well fitted by a single dust component
with temperature $T = 18^{+3}_{-7}$ K 
and emissivity $\epsilon \propto (\nu/\nu_0)^\beta$
with $\beta = 1.9^{+3.0}_{-0.5}$.
The free-free emission is spatially correlated with the dust
on angular scales larger than the 7\deg ~DMR beam,
with {\it rms} variations $5.3 \pm 1.8 ~\mu$K ~at 53 GHz
and angular power spectrum $P \propto \ell^{-3}$.
If this correlation persists to smaller angular scales,
free-free emission should not be a significant contaminant
to measurements of the cosmic microwave anisotropy
at degree angular scales
for frequencies above 20 GHz.

\noindent
{\it Subject headings:} radio continuum: interstellar --
ISM: dust -- cosmic microwave background 

% -------------------- Main text begins on separate page -------------------- 
\clearpage
\section{Introduction}
The microwave sky is dominated by the cosmic microwave background (CMB)
and Galactic synchrotron, dust, and free-free emission.
A number of authors have attempted to distinguish these components
based on 
sky surveys where one component dominates the others
and on 
the frequency spectrum inferred from microphysical processes
(Fixsen, Cheng, \& Wilkinson 1983;
Lubin et al.\ 1985;
Wright et al.\ 1991;
Bennett et al.\ 1992;
Bensadoun et al.\ 1993;
Guti\'{e}rrez de la Cruz et al.\ 1995).
This approach has been most successful for synchrotron radiation:
radio surveys and 
local measurements of the cosmic ray electron energy spectrum 
determine both the spectrum and morphology within broad limits
(Banday \& Wolfendale 1991, Bennett et al.\ 1992).
Free-free emission from electron-ion interactions
has well-determined spectral behavior
but is poorly mapped at high Galactic latitudes
(Bennett et al.\ 1992, Reynolds 1992).
Dust emission suffers the opposite problem: 
although it dominates infrared surveys,
its spectral behavior depends on the shape, composition, and size distribution 
of the dust grains, which are poorly known
(D\'{e}sert, Boulanger, \& Puget 1990).

At millimeter wavelengths, 
the large angular scale anisotropy in the CMB 
is larger than fluctuations in the combined Galactic foregrounds,
so that detailed Galactic modelling is not required for cosmological tests
(Bennett et al.\ 1992, Smoot et al.\ 1992).
At finer angular resolution or across a broader frequency band,
this is not necessarily true
(Banday \& Wolfendale 1990, 1991;
Franceschini et al.\ 1989).
Precise maps of Galactic free-free and dust emission
provide information on physical processes in the interstellar medium
and aid planning of future CMB anisotropy measurements.

Synchrotron emission results from the acceleration of cosmic-ray electrons
in the Galactic magnetic field,
and may be approximated as a power law in frequency,
$ T_{\rm synch} \propto \nu^\beta$.
Bennett et al.\ (1992) review several models of synchrotron emission
based on radio surveys 
at 408 MHz (Haslam et al.\ 1981) 
and 1420 MHz (Reich \& Reich 1988).
They consider three models for the spectral index:
spatially invariant,
spatially varying but frequency-independent,
and a spatially varying index that steepens with frequency
as determined by the local cosmic-ray energy spectrum.
Comparison with a survey at 19 GHz (Boughn et al.\ 1992)
suggests that the third model is somewhat better than the other two,
although all have obvious shortcomings.
Synchrotron indices used by various authors
to scale the 408 MHz survey to millimeter wavelengths
range from -2.7 to -3.1,
corresponding to a factor of 7 uncertainty in amplitude at 53 GHz.

Bennett et al.\ (1992, 1994a) provide a full-sky map of the Galactic free-free 
emission derived from linear combinations of the
{\it COBE} Differential Microwave Radiometers (DMR) sky maps,
designed to isolate emission with $T_{\rm ff} \propto \nu^{-2.15}$.
Although in principle this map presents an unbiased full-sky template
for emission from the ionized gas in the interstellar medium,
in practice the noise levels are such that only the quadrupolar component
$\Delta T \propto \csc|b|$ is detected at $|b| \gt 15\deg$.
Other tracers of the warm ionized medium
suffer from signal to noise ratio
(e.g., N{\tenrm II} emission, Bennett et al.\ 1994b)
or from undersampling and selection biases 
(e.g., H$\alpha$ emission and pulsar dispersion measure, 
Reynolds 1992, 1984).

Wright et al.\ (1991) 
provide estimates of far-infrared Galactic dust emission
using the {\it COBE} Far Infrared Spectrophotometer (FIRAS).
Assuming that the dust intensity along each line of sight 
may be characterized by optically thin emission 
with frequency-dependent emissivity 
$ I_\nu = \tau_0 ~(\frac{\nu}{\nu_0})^\beta ~B_\nu(T) $
where $B_\nu$ is the Planck function 
and $\nu_0 = 900$ GHz,
they find temperature $T = 23.3$ K and emissivity $\beta = 1.65$
for the mean spectrum including the Galactic plane.  
A fit with $\beta$ fixed at 2 shows excess emission at long wavelengths,
characterized by a two-temperature model with 
$T_1 \approx 20.4$ K and $T_2 \approx 4.8$ K.
Reach et al.\ (1995a) extend this model with an improved data set;
at high latitudes $(|b| \gt 30\deg)$ 
they detect a pervasive long wavelength excess 
characterized by
either a flattened emissivity $\beta = 1.2 \pm 0.3$
or a combination of warm and cold components with
$T_1 = 17.6 \pm 0.5$ K and $T_2 = 6.9 \pm 0.8$ K.
The opacities of the warm and cold components are highly correlated;
however, the FIRAS data do not distinguish between two distinct
(although spatially correlated) dust populations 
or a single dust population with enhanced submillimeter emissivity.

Attempts to measure Galactic emission at millimeter wavelengths
are hindered by the small amplitude of the individual emission components
and the presence both of other Galactic sources 
and of anisotropy in the cosmic microwave background.
Point-by-point decomposition based on the different frequency dependences
of the emission mechanisms are typically limited by signal to noise
considerations (Brandt et al. 1994).
The ability to detect weak signals is significantly enhanced if the
angular distribution is known {\it a priori}.  
In this paper,
we cross-correlate the DMR full-sky maps with spatial templates 
from long-wavelength radio surveys
and the far-infrared maps of the 
{\it COBE} Diffuse Infrared Background Experiment 
(DIRBE, Boggess et al.\ 1992).
After accounting for instrument noise 
and chance alignments of the CMB anisotropy,
we derive upper limits to the synchrotron spectral index 
steeper than used by many authors.
We obtain a statistically significant cross-correlation with the DIRBE maps
whose dependence on the DMR frequencies 
indicates a superposition of dust and free-free emission.
The dust temperature and emissivity derived from 
angular variations in the DMR and DIRBE data agree with the values
derived independently from the absolute FIRAS spectra,
and place a lower limit on enhanced dust emissivity at long wavelengths.
The detection of free-free emission correlated with the DIRBE dust emission
provides a spatial template and normalization 
for at least one component 
of the warm ionized medium 
at various angular scales.

% The preprint looks better if this section starts a new page
\clearpage
\section{Analysis}
We assume that the DMR microwave maps may be represented by a superposition
of CMB anisotropy and Galactic signals whose angular distribution is
traced by an external map,
$$
\Delta T^{\rm DMR} =  \Delta T^{CMB} ~+ ~\alpha \Delta X^{\rm Gal},
$$
where $\Delta T^{\rm DMR}$ is the antenna temperature\footnotemark[7]
\footnotetext[7]{
~Antenna temperature $T_A$ is defined in terms of the power received per unit
bandwidth, $P = k T_A \Delta \nu$ where $k$ is Boltzmann's constant.
It is related to the intensity $I_\nu$ by
$I_\nu = 2 k T_A \frac{\nu^2}{c^2}$.
}
in a DMR map (in units mK),
$\Delta X^{\rm Gal}$ is the intensity of the Galactic template map
(not necessarily in temperature units),
and the coefficient $\alpha$ converts the units of the Galactic map
to antenna temperature at the DMR frequency.
Fitting for $\alpha$ on a pixel-by-pixel basis 
(e.g., a linear correlation coefficient)
does not make full use of the 
angular information in the two maps;
instead, we compare the full cross-correlation function
$$
C(\theta) = \sum_{i,j} ~\Delta X^{\rm Gal}_i ~\Delta T^{\rm DMR}_j
$$
(summed over all pixel pairs $\{i,j\}$ separated by angle $\theta$)
to the autocorrelation function
$$
A(\theta) = \sum_{i,j} ~\Delta X^{\rm Gal}_i ~\Delta X^{\rm Gal}_j
$$
and find the coefficient $\alpha$ by minimizing
$$
\chi^2 = \sum_{a,b} (C - \alpha A)_a ({\bf M}^{-1})_{ab} (C - \alpha A)_b
$$
where ${\bf M}$ is the covariance matrix of the cross-correlation function,
and the indices $a$ and $b$ run over the angular bins of the correlation
functions.
Bennett et al.\ (1993) use the same technique to search for
correlation between the DMR maps and various extragalactic sources.

We use the 2-year DMR sky maps (Bennett et al.\ 1994a)
at 9.5, 5.7, and 3.3 mm wavelength (31.5, 53, and 90 GHz).
For long wavelength (synchrotron) Galactic templates, we use either the
408 MHz survey (Haslam et al.\ 1981)
or the 408 MHz survey scaled to the DMR frequencies using
a spatially varying frequency-dependent spectral index
as described in Bennett et al.\ (1992),
denoted here as the ``cosmic ray'' model.
For short wavelength (dust) Galactic templates, we use the 
DIRBE sky maps at 240, 140, and 100 $\mu$m wavelength
from which a model of the zodiacal dust emission 
has been removed (Reach et al.\ 1995b).
We convolve the Galactic template maps 
with the DMR beam pattern (Wright et al.\ 1994) 
and pixelize them in the same Galactic representation used for the DMR maps.

The Galactic plane has considerable structure 
smaller than the 7\deg ~DMR beam size;
we exclude the Galactic plane ($|b| \lt 20\deg$ or $|b| \lt 30\deg$)
from all analysis.
The DMR maps are insensitive to a monopole
and have a dipole dominated by the CMB.
Galactic microwave emission from synchrotron, free-free, and dust 
is dominated by a csc$|b|$ quadrupole.
Analysis which includes the quadrupole 
could result in a false positive cross-correlation 
between the DMR maps and a Galactic template
due to the microwave emission from a different template
with a brighter quadrupole at millimeter wavelengths.
Furthermore, the CMB quadrupole has a component counter-aligned 
with a Galactic cosecant of similar magnitude (Bennett et al.\ 1992),
potentially masking correlation between the DMR and Galactic maps.
Accordingly, we remove the monopole, dipole, and quadrupole from the
the high-latitude portion of both the DMR and Galactic template maps
prior to computing the auto- and cross-correlation functions.
As a consequence, our results depend only on 
the intensity variations in the maps
and are insensitive to the zero level 
of either the DMR or Galactic template maps.

We determine the covariance matrix ${\bf M}$ of the cross-correlation function
and assess the statistical significance of the coefficient $\alpha$
as Galactic features occasionally align with CMB anisotropy
by using Monte Carlo simulations in which the DMR map is replaced by
a scale-invariant Harrison-Zel'dovich spectrum of CMB anisotropy normalized
to the quadrupole value 
$Q_{rms-PS} = 20 ~\muK$ determined from the 2-year DMR data
(G\'{o}rski et al.\ 1994).  
To each CMB realization we add a realization of instrument noise
defined by the level and pattern of noise in the DMR A and B channels,
then combine the channels to form the (A+B)/2 sum map.
We excise the Galactic plane and remove the monopole, dipole, and quadrupole
from each simulated sum map prior to computing the cross-correlation.
The mean coupling between angular bins defines the covariance matrix,
$ {\bf M}_{ab} =  \frac{1}{N} \sum ~ C_a C_b $
averaged over $N$ simulations;
throughout this work we use 72 angular bins each of width 2\ddeg6.
The simulations thus automatically account for
cosmic variance from random alignments 
of CMB anisotropy with features in the Galactic template maps
as well as the effect of instrument noise and
the aliasing of power between multipole moments 
in the quadrupole-removed maps.
Cosmic variance dominates the uncertainty in the fitted $\alpha$;
consequently, we weight each pixel uniformly when computing
the cross-correlation function.

We test for possible biases in the cross-correlation technique 
by analyzing 2000 additional simulations which include
a signal $T^{\rm sim} ~= ~0.002 
~{\rm mK} ~{\rm MJy}^{-1} ~{\rm sr} ~I^{\rm DIRBE}$
in addition to the CMB and noise.
We recover a mean coefficient 
$\langle \alpha \rangle = 
(1.96 \pm 0.03) \times 10^{-3}
~{\rm mK} ~{\rm MJy}^{-1} ~{\rm sr}$
at $|b| > 20\deg$,
in excellent agreement with the simulation input.
We have also repeated the analysis using
the cross power spectra 
and obtained coefficients $\alpha$
in agreement with those found using the cross correlation function $C(\theta)$.

% The preprint looks better if this section starts a new page
\clearpage
\section{Results}

Table 1 shows the fitted coefficient $\alpha$ from the 
cross correlation of the DMR 2-year maps 
with various Galactic template maps,
compared to 2000 random realizations of CMB and instrument noise.  
We detect no statistically significant correlation between the DMR maps
and any synchrotron template: the amplitude of any signal with the
the 408 MHz survey or the cosmic-ray synchrotron model
is smaller than the noise in the DMR maps.
All of the DMR maps show a statistically significant correlation 
with the DIRBE far-infrared maps,
indicating a detection of signal with common spatial structure
in the high-latitude portion of the two data sets.

The correlation coefficients are averages over the high-latitude sky and
do not provide specific information on individual patches of the sky.
We estimate the amplitude of the signal at each DMR frequency
by multiplying the standard deviation of the Galactic template maps
(after quadrupole subtraction) by the fitted coefficients $\alpha$.
Since the three DIRBE maps trace the same warm dust emission,
the DIRBE templates provide nearly identical estimates 
of Galactic emission at each of the DMR frequencies.
The uncertainties in the fitted coefficients $\alpha$ 
between any one DMR map and the three DIRBE maps
are dominated by the noise and CMB signal in the DMR maps,
and are not statistically independent.
We therefore adopt the unweighted average of the fitted signal
at each DMR frequency,
with uncertainty estimated by the smallest uncertainty
of the three DIRBE cross-correlations.

Table 2 shows the estimated Galactic signal in the DMR maps
traced by the two synchrotron templates and the DIRBE dust morphology.
The Galactic signal correlated with the DIRBE dust distribution
has {\it rms} fluctuations 
$\Delta T_{\rm Gal} = 7.1 \pm 3.4 ~\mu$K antenna temperature
at $|b| > 30\deg$ and 7\deg ~angular resolution
in the most sensitive 53 GHz channels.
For comparison, the {\it rms} instrument noise at 53 GHz is 98 $\mu$K.
Galactic emission is a minor contribution to the brightness fluctuations
at millimeter wavelengths.
Specifying the template of the Galactic emission 
allows a significant increase in sensitivity to faint extended emission.

Since the DIRBE maps are dominated by infrared dust emission,
it is natural to interpret the correlated DMR--DIRBE signal as evidence for
dust emission at millimeter wavelengths.
However, the sharp rise in signal amplitude at 53 and 31.5 GHz
is inconsistent with emission by dust alone.
The amplitude (in antenna temperature) of emission traced by the DIRBE maps
more than doubles between 53 and 31.5 GHz,
requiring some emission component with negative spectral index.
The spectral index $T \propto \nu^\beta$ of the signal,
without modelling any components, is
$\beta = -1.7 \pm 0.4$ between the 31.5 and 53 GHz data and
$\beta = -1.5 \pm 0.9$ between the 53 and 90 GHz data (68\% confidence),
suggestive of an admixture of dust ($\beta \approx +2$) 
and free-free ($\beta = -2.15$) emission,
but inconsistent with dust emission only.

The DMR--DIRBE signals are not appreciably affected by synchrotron emission
($\beta_{\rm synch} \approx -3.1$ at millimeter wavelengths).
Direct cross-correlation of the DMR maps with 
either the 408 MHz survey or the cosmic-ray synchrotron model
shows no statistically significant signal.
The fitted coefficients for the 408 MHz survey at $|b| > 20\deg$
are significant only at 90\% confidence,
and have a spectrum closer to the CMB than to synchrotron emission (Table 2).
The 408 MHz coefficients all become negative 
at the higher cut $|b| > 30\deg$,
which we take as further evidence that the 
correlations at $|b| > 20\deg$ are dominated by 
chance alignment of the CMB with a feature in the 408 MHz survey.
Similarly, there is no compelling evidence
for synchrotron emission in the 2-year DMR maps
with spatial template traced by the cosmic-ray model.
We adopt an upper limit for either model
$\Delta T_{\rm synch} < 12 ~\mu$K (95\% confidence) 
for fluctuations in synchrotron emission at 31.5 GHz.

Cross-correlations between the DIRBE maps and the 408 MHz survey
provide further evidence that 
the rising signal in the low-frequency DMR channels is not
caused by synchrotron emission.
We cross-correlate the DIRBE and 408 MHz template maps
(with the 408 MHz survey scaled to 53 GHz 
using a spectral index $\beta = -3$ to provide
direct comparison with the DMR 53 GHz results)
and obtain a fitted coefficient at $|b| > 30\deg$,
$\alpha_{\rm synch} = 3 \times 10^{-5}$ mK MJy$^{-1}$ sr,
two orders of magnitude smaller than the correlation observed between
the DMR and DIRBE maps.
Synchrotron emission at high latitude
does not strongly correlate with the DIRBE far-infrared dust emission.

We conclude that synchrotron emission does not account for the observed
correlation between the high-latitude DMR and DIRBE sky maps,
leaving free-free emission as the most likely alternative.
We decompose the sky signals from Table 2 
into dust and free-free components 
assuming $\beta=2$ for the dust.
Table 3 shows the {\it rms} dust and free-free normalization at 53 GHz.
Approximately half of the high-latitude Galactic signal 
in the DMR 90 GHz data results from dust and half from free-free emission,
while the 53 GHz data are dominated by free-free emission.
Free-free emission increases markedly at the lower Galactic cut.
Note that $|b| > 20\deg$ includes emission 
from the Orion and Ophiuchus complexes,
which contain substantial amounts of ionized gas.
The increase in fitted free-free amplitude when regions 
known to contain ionized gas are included in the analysis
provides further support for the existence of 
free-free emission spatially correlated with the infrared dust emission.

Tables 1 and 2 demonstrate a significant detection of dust and free-free 
emission in the high-latitude DMR data.
We may use the combined DIRBE and DMR data to limit 
dust emission properties at long wavelengths.
We fit the {\it rms} DIRBE and DMR signals to dust models of the form
$I_\nu = \tau (\frac{\nu}{\nu_0})^{\beta} B_\nu(T) 
	~+ A_{\rm ff}(\frac{\nu}{\nu_0})^{-0.15}$ (model 1) or
$I_\nu = \tau_1 (\frac{\nu}{\nu_0})^2 B_\nu(T_1) ~+
	~\tau_2 (\frac{\nu}{\nu_0})^2 B_\nu(T_2)
	~+ A_{\rm ff}(\frac{\nu}{\nu_0})^{-0.15}$ (model 2),
i.e., a model with a single dust population 
with enhanced submillimeter emissivity
or a two-temperature model with $\nu^2$ emissivity.
Both models include a free-free emission term $A_{\rm ff}$.
Figure 1 shows the spectra of the intensity fluctuations
and the fitted dust and free-free emission 
for a single dust component (model 1) at $|b| > 20\deg$.
Figure 2 shows the $\chi^2$ contours of this model in the
temperature-emissivity plane.
The best fit occurs for
dust temperature $T = 18^{+3}_{-7}$ K and $\beta = 1.9^{+3.0}_{-0.5}$
with $\tau = (1.2^{+2.5}_{-0.5}) \times 10^{-5}$ (68\% confidence),
in agreement with the absolute FIRAS spectra 
(Wright et al.\ 1991, Reach et al.\ 1995a).
The DMR data place a lower limit 
to a frequency-dependent dust emissivity
over the wavelength range 6 mm to 100 $\mu$m:
$\beta \gt 1.2$ at 95\% confidence.

The two-component dust model is not significantly constrained 
by the DMR-DIRBE cross-correlation.  
Provided one component is near 18 K, 
changes in the temperature of the cold component
are compensated by correlated variations in opacity.
We may fix the temperatures in model 2 to the values derived
from the FIRAS $|b| \gt 30\deg$ spectra
and compare the opacities 
derived from fluctuations in the warm and cold components 
to the total opacities derived from FIRAS (Table 4).
The opacities for both the warm and cold components 
derived from the DMR-DIRBE cross-correlation are smaller
than the corresponding values derived from the FIRAS spectra,
as expected if the dust emission 
contains a significant monopole or quadrupole term.

Figure 3 summarizes the contribution of various Galactic emission mechanisms
(standard deviation of each component averaged over $|b| \gt 30 \deg$) 
at 7\deg ~angular resolution.
The width of each band reflects the 68\% confidence level uncertainty in the
amplitude of each component, and includes the absolute calibration
uncertainty of the template maps.  
The ``cosmic-ray'' synchrotron model 
provides a modestly better fit to the available low-frequency data
than the 408 MHz survey scaled with a spatially invariant spectral index
(Bennett et al.\ 1992).
Accordingly, we normalize the synchrotron curve in Figure 3 
by the cross-correlation of the DMR 31.5 GHz channel with the cosmic-ray model,
as indicated by the filled circle.  
Fluctuations in the synchrotron emission at DMR frequencies 
are modestly smaller than expected;
the cosmic-ray model predicts $\Delta T_{\rm synch} = 13$ \muK
~at $|b| \gt 30\deg$ in the 31.5 GHz map.
We note that our results are insensitive to the zero level 
of the radio surveys, 
which can influence the amplitude of the predicted fluctuations
through a monopole change in the spectral index.
The 95\% confidence level upper limit on synchrotron emission
($\Delta T_{\rm synch} \lt 12 ~\muK$ ~at 31.5 GHz) is 
the same for both the cosmic-ray and 408 MHz templates,
and may be used to set an upper limit on 
the high-latitude synchrotron spectral index:
$\beta_{\rm synch} \lt -2.9$
between 408 MHz and 31.5 GHz.

Free-free emission is normalized by the free-free values in Table 2
and dominates fluctuations in diffuse synchrotron emission for frequencies
above 20 GHz.
The dust spectrum inferred from the DMR--DIRBE cross-correlation is in 
agreement with more precise spectra from the {\it COBE} FIRAS experiment.
The dust model in Figure 3 represents
a range of high-latitude spectra from
$T_{\rm dust} = 21$ K and $\beta_{\rm dust} = 1.4$
to
$T_{\rm dust} = 16$ K and $\beta_{\rm dust} = 2$,
with opacity normalized to the spatial variation of the DIRBE 240 $\mu$m map.
The upper limits on dust emission at millimeter wavelengths
from the DMR--DIRBE cross-correlation dust model
(after removing the free-free component)
are also shown.

% The preprint looks better if this section starts a new page
\clearpage
\section{Discussion}
A significant result of this analysis 
is the detection of fluctuations in the microwave sky brightness 
with the angular variation of the DIRBE far-infrared maps 
and the frequency spectrum of free-free emission.
The angular structure in the far-infrared maps 
is dominated by interstellar dust associated with atomic gas
(Boulanger \& P\'{e}rault 1987).
Our results suggest that the ionized component of the interstellar medium
is at least partially correlated with the atomic gas.

Previous comparisons of the H$\alpha$ recombination line from the ionized gas 
revealed no clear correlation
with the H{\tenrm I} hyperfine line from the neutral gas (Reynolds 1987),
although one filament detected in H$\alpha$ may be related 
to a parallel offset structure in H{\tenrm I} (Reynolds et al.\ 1995).
Similar correlated structures, 
as well as evaporating surfaces of clouds 
embedded in a hot ionized gas
({\it cf} McKee \& Ostriker 1977)
could account for the correlation we observe
between the dust and free-free emission.
The {\it rms} variation in the inferred free-free emission at 53 GHz is
$\Delta T_{\rm ff} = 5.3 \pm 1.8 ~\mu$K
at 7\deg ~angular resolution,
averaged over the high-latitude sky ($|b| \gt 30\deg$)
after removal of the quadrupole component.
The amplitude of this signal compares favorably
with the predicted signal
$\Delta T_{\rm ff} \approx 7 ~\muK$
inferred from fluctuations in H$\alpha$ emission (Reynolds 1992),
where for convenience we reference $\Delta T_{\rm ff}$ to 53 GHz.

The detection of free-free emission 
correlated with the DIRBE far-infrared maps
provides a template for at least one component of Galactic free-free emission.
We may place an upper limit for the combined free-free emission 
from all sources by forming a linear combination of the DMR maps
designed to pass emission with spectral index -2.15,
cancel emission with a CMB spectrum,
and minimize instrument noise:
$$
T_{\rm ff} = 
  0.40 \times \frac{1}{2}(T^\prime_{\rm 31A} \pm T^\prime_{\rm 31B}) 
- 0.09 \times \frac{1}{2}(T^\prime_{\rm 53A} \pm T^\prime_{\rm 53B}) 
- 0.37 \times \frac{1}{2}(T^\prime_{\rm 90A} \pm T^\prime_{\rm 90B}),
$$
where $T^\prime$ is the antenna temperature in each DMR channel
after subtracting synchrotron and dust emission 
using the cosmic-ray and DIRBE models, respectively.
Since both synchrotron and dust are small compared to the free-free emission
at 53 GHz (Figure 3), 
their residual contribution after correction is negligible.
This linear combination has only 33\% larger noise than the most 
sensitive 53 GHz channels.
We smooth the maps with a 7\deg ~FWHM Gaussian 
to reduce further the effects of noise,
remove a fitted quadrupole, and
compare the variance of the (A+B)/2 sum map to the (A-B)/2 difference map
to obtain an estimate for the free-free fluctuations
$\Delta T_{\rm ff} = 3.9 \pm 5.7$ \muK,
in excellent agreement with the value
$\Delta T_{\rm ff} = 5.1 \pm 1.6$ \muK
~from the DMR-DIRBE cross-correlation 
at the same 10\deg ~effective smoothing.
Free-free emission from sources uncorrelated with the far-infrared dust 
emission, although present at high latitudes,
does not dominate the component correlated with the dust emission
for the range of angular scales probed here.

Figure 4a shows the power spectrum 
of the inferred free-free emission at 53 GHz.
We calculate the power spectrum of the high-latitude DIRBE 240 $\mu$m map
using a set of basis functions 
orthonormal on the region $|b| \gt 30\deg$
(G\'{o}rski 1994),
and normalize to free-free emission at 53 GHz using the
correlation coefficient derived from a 
dust/free-free decomposition of the correlation coefficients from Table 1,
$\alpha_{\rm ff} = (3.35 \pm 1.11) \times 10^{-3}$ mK MJy$^{-1}$ sr.
The power falls rapidly at small angular scales:
$$
P_{\rm ff} = \frac{8 a_2}{\ell^{3}},
$$
where
$P(\ell) = \frac{1}{2\ell + 1} \sum_m a_{\ell m}^2$
is the mean power at multipole order $\ell$
($\ell ~\approx ~\pi/\theta$),
and the quadrupole normalization $a_2 = (6.02 \pm 3.99) ~\mu{\rm K}^2$.
The power spectrum in Figure 4 is an average over the high-latitude sky.
On smaller angular scales,
Gautier et al.\ (1992) estimate the dust power spectrum
using $8\deg \times 8\deg$ IRAS patches
and find $P \propto \ell^{-3}$ as well.

The DMR and DIRBE data provide the normalization
for the correlated free-free emission on angular scales larger than 7\deg.
The $\ell^{-3}$ dust power spectrum continues to smaller angular scales,
and it is plausible that the spatial correlation 
between the dust and the warm ionized gas
also persists.
We thus estimate the {\it rms} free-free fluctuations 
on various angular scales
by integrating the free-free power spectrum in Figure 4a
over the range in $\ell$ corresponding to the
patch size and the Gaussian beam width.
Table 5 compares the model predictions to
H$\alpha$ and radio measurements.
An H$\alpha$ map of a single $12\deg \times 10\deg$ patch 
suggests {\it rms} variations 
$\Delta T_{\rm ff} = 1.4 ~\muK$
at 0\ddeg8 angular resolution (Reynolds 1992).
Similar observations of a $7\deg \times 7\deg$ patch
at 0\ddeg1 angular resolution yield an upper limit of
$\Delta T_{\rm ff} \lt 1.2 ~\muK$
~(Gaustad et al.\ 1995),
while deep VLA observations at 1\amin resolution
provide an upper limit
$\Delta T_{\rm ff} \lt 1.3 ~\muK$ 
~(Fomalont et al.\ 1993).
There is generally good agreement
between observations and the power spectrum predictions,
suggesting that 
the dominant free-free emission is correlated with the dust
and that the correlation persists to smaller angular scales.

If the free-free/dust correlation does persist 
to smaller angular scales,
it has interesting implications 
for experiments measuring CMB anisotropy at degree angular scales.
Measurements of CMB anisotropy must contend with the problem of
estimating the foreground Galactic emission 
and removing it if it contributes a significant fraction of the CMB anisotropy.
Since no template for the free-free emission exists,
experimenters typically observe in several frequency bands
and use a linear combination 
of the angular anisotropy at different frequencies
to remove any emission with the free-free emission spectrum.
Although this removes the foreground emission,
it increases the noise in the corrected CMB results,
often to the point of obscuring the CMB signal.
If free-free emission could be demonstrated 
to be a negligible contaminant,
significant reductions in noise could be realized.

Table 5 shows the free-free power spectrum predictions for
two cases of interest.
A full-sky map at 53 GHz with angular resolution 0\ddeg5
would observe {\it rms} free-free emission of $6.1 \pm 2.0$ \muK
~averaged over the high-latitude sky
after quadrupole subtraction.
Surveys restricted to smaller regions exclude power from
Fourier modes with wavelength larger than the survey size.
A $10\deg \times 10\deg$ survey 
(corresponding to modes $\ell \gt 18$)
with angular resolution 0\ddeg5
would measure {\it rms} free-free emission of $1.5 \pm 0.5$ \muK,
where the quoted errors 
reflect only the uncertainty in the free-free normalization
and do not include the variation from patch to patch on the sky.
Emission at this level is well below the expected magnitude
of CMB anisotropy,
although individual patches may show free-free emission significantly larger
than the high-latitude average.

The small amplitude of the free-free signal
explains why no structure is observed in the
maps of free-free emission
derived from linear combinations of the two-year DMR maps.
The instrument noise in the free-free map presented above
is 20 times larger than the free-free signal estimated from
the DMR--DIRBE cross-correlation.
Figure 4b shows the power spectrum of this free-free map,
including the uncertainty from instrument noise.
It is clear that the noise per multipole order $\ell$ 
is much larger than the signal.
The increase in sensitivity required to detect signals 
at the few \muK ~level
results from the {\it a priori} specification of the DIRBE map
as a spatial template for the emission.

The high-latitude Galactic signal 
is small compared to the CMB anisotropy in the DMR maps
and does not significantly alter previously published estimates
based on the uncorrected DMR 53 GHz sky maps.
The standard deviation 
at 7\deg ~angular resolution
of the combined CMB and Galactic signals 
at $|b| \gt 30\deg$ is
$\Delta T_{\rm sky} = 42.6 \pm 8.4 ~\muK$
antenna temperature at 53 GHz.
The CMB and Galactic signals are uncorrelated;
correcting $\Delta T_{\rm sky}$ for the
estimated Galactic signal $\Delta T_{\rm Gal} = 7.1 \pm 3.4 ~\muK$
yields an estimated CMB anisotropy
$\Delta T_{\rm CMB} = 42.0 \pm 8.6 ~\muK$,
within 0.7 \muK ~(7\% of the statistical uncertainty)
of the uncorrected result.
The CMB anisotropy may also be characterized by 
a power-law distribution of primordial density perturbations
$P(k) ~\propto ~k^n$
with quadrupole normalization $Q_{rms-PS}$ and index $n$.
Removing the Galactic emission traced by the DIRBE maps
changes $Q_{rms-PS}$ or $n$ by less than 0.3 standard deviations,
well within their respective statistical uncertainties.

\section{Conclusions}

We detect statistically significant cross-correlation
between the {\it COBE} two-year DMR maps
and the DIRBE 240, 140, and 100 $\mu$m maps.
The {\it rms} amplitude of the 
correlated Galactic signal in the DMR maps at $|b| > 30\deg$,
after quadrupole subtraction,
is $15.6 \pm 6.0, ~7.1 \pm 3.4$, and $3.9 \pm 3.8 ~\mu$K
at 31.5, 53, and 90 GHz,
corresponding to dust emission 
$\Delta T_{\rm dust} = 0.9 \pm 1.3 ~\mu$K
and free-free emission
$\Delta T_{\rm ff} = 5.3 \pm 1.8 ~\mu$K
at 53 GHz and 7\deg ~angular resolution.
The dust emission may be used to put a lower limit on
the emissivity at millimeter wavelengths,
$\Delta T_{\rm dust} \propto \nu^\beta$ 
with $\beta > 1.2$ at 95\% confidence.
The dust temperature and emissivity derived from fluctuations 
in the DMR and DIRBE maps are in agreement with values
derived from the absolute FIRAS spectra at wavelengths
between the DMR and DIRBE coverage.
Similar analyses using the 408 MHz survey
or a model of synchrotron emission based on the 408 MHz survey
with spatially varying, frequency-dependent spectral index
yielded upper limits for fluctuations in synchrotron emission of
$\Delta T_{\rm synch} < 12 ~\mu$K
at 31.5 GHz,
corresponding to a spectral index
$\beta_{\rm synch} < -2.9$ between 408 MHz and 31.5 GHz.

We detect a component of Galactic free-free emission
correlated with the DIRBE maps,
providing a template for the angular distribution
of free-free emission on angular scales larger than the 7\deg ~DMR beam.
The {\it rms} variations in this correlated component,
$\Delta T_{\rm ff} = 5.3 \pm 1.8 ~\mu$K,
are in good agreement with the {\it rms} variations from {\it all} sources
of free-free emission,
$\Delta T_{\rm ff} = 3.9 \pm 5.7 ~\mu$K,
derived from a linear combination of the DMR 2-year maps.
Extrapolating the power spectrum of the free-free emission
to smaller angular scales yields an estimated contribution
for {\it rms} fluctuations at 0\ddeg5 angular scale
observed at 53 GHz of
$\Delta T_{\rm ff} = 6.1 \pm 2.0 ~\mu$K
for a full-sky map at $|b| > 30\deg$,
or 
$\Delta T_{\rm ff} = 1.5 \pm 0.5 ~\mu$K for observations
restricted to a $10\deg \times 10\deg$ patch.

If the spatial correlation between the dust and warm ionized gas 
observed on large angular scales
persists to smaller angular scales,
free-free emission should not be a serious contaminant
to measurements of the medium-scale CMB anisotropy.
It is worth pointing out, however, 
that small regions of the sky have variations much larger than
the standard deviation over the full sky.
If free-free emission is correlated with dust emission
on all angular scales of interest,
the IRAS or DIRBE maps may be used to estimate both dust
and free-free emission in individual regions of the sky.

% -------------------- References -------------------- 
% Put references on a separate page
\clearpage
\begin{center}
\large
{\bf References}
\end{center}

\refitem
Banday, A.\ \& Wolfendale, A.W.\ 1991, MNRAS, 248, 705

\refitem
--- 1990, MNRAS, 245, 182

\refitem
Bennett, C.L., et al.\ 1994a, ApJ, 436, 423

\refitem
--- 1994b, ApJ, 434, 587

\refitem
--- 1993, ApJ, 414, L77

\refitem
--- 1992, ApJ, 396, L7

\refitem
Bensadoun, M., Bersanelli, M., De Amici, G., Kogut, A., Levin, S.M., Limon, M., 
Smoot, G.F., \& Witebsky, C.\ 1993, ApJ, 409, 1

\refitem
Boggess, N.W., et al.\ 1992, ApJ, 397, 420

\refitem
Boughn, S.P., Cheng, E.S., Cottingham, D.A., \& Fixsen, D.J., 
1992, ApJ, 391, L49

\refitem
Boulanger, F.\ \& P\'{e}rault, M.\ 1987, ApJ, 330, 964

\refitem
Brandt, W.N., Lawrence, C.R., Readhead, A.C.S., 
Pakianathan, J.N., \& Fiola, T.M.
1994, ApJ, 424, 1

\refitem
D\'{e}sert, F.-X., Boulanger, F., \& Puget, J.-L.\ 1990, A\&A, 327, 215

\refitem
Fomalont, E.B., Partridge, R.B., Windhorst, R.A, \& Lowenthal, J.D.
1993, ApJ, 404, 8

\refitem
Fixsen, D.J., Cheng, E.S., \& Wilkinson, D.T.\ 1983, PRL, 50, 620

\refitem
Franceschini, A., Toffolatti, L., Danese, L., \& De Zotti, G.\ 1989,
ApJ, 344, 35

\refitem
Gaustad, J.E., Oh, E.S., McCullough, P.R., \& Van Buren, D.\ 1995,
BAAS, 27, 823

\refitem
Gautier, T.N., Boulanger, F., P\'{e}rault, M., \& Puget, J.L.\ 1992,
AJ, 103, 1313

\refitem
G\'{o}rski, K.M.\ 1994, ApJ, 430, L85

\refitem
G\'{o}rski, K.M., Hinshaw, G., Banday, A.J., Bennett, C.L., Wright, E.L.,
Kogut, A., Smoot, G.F., \& Lubin, P.\ 1994, ApJ, 430, L89

\refitem
Guti\'{e}rrez de la Cruz, C.M., Davies, R.D., Rebolo, R., Watson, R.A., 
Hancock, S., \& Lasenby, A.N.\ 1995, ApJ, 442, 10

\refitem
Haslam, C.G.T, Klein, U., Salter, C.J., Stoffel, H, Wilson, W.E., Cleary, M.N., 
Cooke, D.J., \& Thomasson, P.\ 1981, A\&A, 100, 209

\refitem
Lubin, P., Villela, T., Epstein, G., \& Smoot, G.\ 1985, ApJ, 298, L1

\refitem
McKee, C.F.\ \& Ostriker, J.P.\ 1977, ApJ, 218, 148

\refitem
Reach, W.T., et al.\ 1995a, ApJ, 451, 188

\refitem
Reach, W.T., Franz, B.A., Kelsall, T., \& Weiland, J.L.\ 1995b, 
{\it Unveiling the Cosmic Infrared Background}, ed.\ E.\ Dwek, (New York:AIP)

\refitem
Reich, P., \& Reich, W.\ 1988, A\&AS, 74, 7

\refitem
Reynolds, R.J., Tufte, S.L., Kung, D.T., McCullough, P.R., \& Heiles, C.\
1995, ApJ, (submitted)

\refitem
Reynolds, R.J.\ 1992, ApJ, 392, L35

\refitem
--- 1987, ApJ, 323, 118

\refitem
--- 1984, ApJ, 282, 191

\refitem
Smoot, G.F., et al.\ 1992, ApJ, 396, L1

\refitem
Wright, E.L., et al.\ 1994, ApJ, 420, 1

\refitem
--- 1991, ApJ, 381, 200

% --------------------------------------------------------------------
% Tables go on a separate page
% --------------------------------------------------------------------
%
% Table 1: Correlation coefficient between the raw sky maps
\vfill
\clearpage

\normalsize
\halfspace
\begin{table}
\caption{DMR-Galactic Template Cross-Correlation Coefficients$^a$}
\begin{center}
\begin{tabular}{l r c c}
\hline
Template Map & DMR Map & $10^3 ~\alpha ~(|b| \gt 20\deg)$ & 
$10^3 ~\alpha ~(|b| \gt 30\deg)$ \\
\hline
408 MHz$^b$          & 31.5 GHz &  2.32 $\pm$ 1.57 & -0.70 $\pm$ 2.25 \\
                     & 53 GHz   &  1.54 $\pm$ 0.91 & -0.28 $\pm$ 1.31 \\
                     & 90 GHz   &  1.14 $\pm$ 1.00 & -0.068 $\pm$ 1.43 \\
 & & & \\
Cosmic-Ray$^c$       & 31.5 GHz &  (0.21 $\pm$ 0.40) $\times ~10^3$ &  (0.63 $\pm$ 0.46) $\times ~10^3$ \\
                     & 53 GHz   &  (0.13 $\pm$ 0.93) $\times ~10^3$ &  (0.78 $\pm$ 1.07) $\times ~10^3$ \\
                     & 90 GHz   &  (3.83 $\pm$ 5.25) $\times ~10^3$ & (-2.24 $\pm$ 5.95) $\times ~10^3$ \\
 & & & \\
DIRBE 240 $\mu$m$^d$ & 31.5 GHz & 12.63 $\pm$ 1.80 & 10.09 $\pm$ 3.65 \\
                     & 53 GHz   &  5.61 $\pm$ 1.03 &  4.03 $\pm$ 2.06 \\
                     & 90 GHz   &  2.34 $\pm$ 1.20 &  2.34 $\pm$ 2.28 \\
 & & & \\
DIRBE 140 $\mu$m$^d$ & 31.5 GHz &  9.27 $\pm$ 1.33 &  7.38 $\pm$ 2.90 \\
                     & 53 GHz   &  4.06 $\pm$ 0.77 &  3.90 $\pm$ 1.70 \\
                     & 90 GHz   &  1.82 $\pm$ 0.86 &  1.47 $\pm$ 1.83 \\
 & & & \\
DIRBE 100 $\mu$m$^d$ & 31.5 GHz & 18.06 $\pm$ 2.54 & 14.57 $\pm$ 6.03 \\
                     & 53 GHz   &  6.88 $\pm$ 1.40 &  6.46 $\pm$ 3.45 \\
                     & 90 GHz   &  2.76 $\pm$ 1.61 &  4.56 $\pm$ 3.89 \\
\hline
\end{tabular}
\end{center}
$^a$ The antenna temperature at each DMR frequency
of Galactic emission traced by template map X
is $T_A = \alpha X$ mK (see text). \\
$^b ~\alpha$ has units mK K$^{-1}$ since the template map has units K.\\
$^c ~\alpha$ is dimensionless since the template map has units mK.\\
$^d ~\alpha$ has units mK (MJy/sr)$^{-1}$ since the template map 
has units MJy sr$^{-1}$.
\end{table}

% -------------- Table 2: Estimated signal at DMR frequencies -------------- 
\normalsize
\halfspace
\begin{table}
\caption{RMS Galactic Signal in DMR Sky Maps}
\begin{center}
\begin{tabular}{l c c c}
\hline
Template Map & DMR Frequency & {\it rms} at $|b| \gt 20\deg$ &
{\it rms} at $|b| \gt 30\deg$ \\
 & (GHz)	&  (\muK) & (\muK) \\
\hline
408 MHz    & 31.5 &  7.6 $\pm$ 5.1  & -1.9 $\pm$ 6.1 \\
           & 53	  &  5.0 $\pm$ 3.0  & -0.8 $\pm$ 3.6 \\
           & 90	  &  3.7 $\pm$ 3.3  & -1.9 $\pm$ 3.9 \\
 & & & \\
Cosmic-Ray & 31.5 &  2.9 $\pm$ 5.4  & ~8.0 $\pm$ 5.8 \\
           & 53   &  0.3 $\pm$ 2.8  & ~2.2 $\pm$ 3.0 \\
           & 90   &  2.4 $\pm$ 3.2  & -1.3 $\pm$ 3.5 \\
 & & & \\
DIRBE      & 31.5 & 35.2 $\pm$ 4.9  & 15.6 $\pm$ 6.0 \\
           & 53   & 14.8 $\pm$ 2.7  & ~7.1 $\pm$ 3.4 \\
           & 90   &  6.3 $\pm$ 3.1  & ~3.9 $\pm$ 3.8 \\
\hline
\end{tabular}
\end{center}
\end{table}

%
% --------------- Table 3: Dust and free-free components ---------------
\normalsize
\halfspace
\begin{table}
\caption{Free-Free and Dust Antenna Temperature at 53 GHz}
\begin{center}
\begin{tabular}{l c c}
\hline
Component & {\it rms} at $|b| \gt 20\deg$ & {\it rms} at $|b| \gt 30\deg$ \\
	& (\muK)	  & (\muK) \\
\hline
Free-Free       & 12.0 $\pm$ 1.5 & 5.3 $\pm$ 1.8 \\
Dust            &  1.1 $\pm$ 1.1 & 0.9 $\pm$ 1.3 \\
\hline
\end{tabular}
\end{center}
\end{table}

% ---------- Table 4: Two-component model for DMR--DIRBE vs FIRAS ----------
\normalsize
\halfspace
\begin{table}
\caption{Dust Opacities in Two-Component Model, $|b| \gt 30\deg$}
\begin{center}
\begin{tabular}{l c c}
\hline
Data Set	& Warm Opacity    & Cold Opacity \\
		& $10^5 ~\tau_1$  & $10^5 ~\tau_2$ \\
\hline
FIRAS$^a$	& 2.18 $\pm$ 0.03 & 15.0 $\pm$ 0.7 \\
DMR-DIRBE$^b$	& 0.87 $\pm$ 0.08 & 2.3$^{+ 11}_{-2.3}$ \\
\hline
\end{tabular}
\end{center}
$^a$~FIRAS values derived from mean dust spectra (Reach et al.\ 1995a). 

$^b$~DMR-DIRBE values derived from intensity fluctuations.
\end{table}

% Table 5: Comparison of free-free observations with model predictions
\normalsize
\halfspace
\begin{table}
\caption{Free-Free Observations and Model Predictions$^a$}
\begin{center}
\begin{tabular}{l l l l l}
\hline
Angular Resolution & Patch Size & Observed $\Delta T_{\rm ff}$ & 
Predicted $\Delta T_{\rm ff}^b$ & Reference \\
\hline
10\deg  & $|b| \gt 30\deg$       & 3.9 $\pm$ 5.7 & 5.3 $\pm$ 1.6 & This work \\
0\ddeg8 & $12\deg \times 10\deg$ & 1.4           & 1.5 $\pm$ 0.5 & Reynolds 1992 \\
0\ddeg1 & $7\deg \times 7\deg$   & $\lt$ 1.2     & 1.4 $\pm$ 0.5 & Gaustad et al. 1995 \\
1\amin  & $7\amin \times 7\amin$ & $\lt$ 1.3     & 0.15 $\pm$ 0.05 & Fomalont et al. 1993 \\
0\ddeg5 & $10\deg \times 10\deg$ & ---           & 1.5 $\pm$ 0.5 & --- \\
0\ddeg5 & $|b| \gt 30\deg$       & ---           & 6.1 $\pm$ 2.0 & --- \\
\hline
\end{tabular}
\end{center}
$^a$~All free-free values are in \muK ~antenna temperature at 53 GHz \\
$^b$~{\it rms} fluctuations estimated from
power spectrum of free-free emission correlated with DIRBE dust emission
(see text).
\end{table}

% --------------------------------------------------------------------
% Encapsulated PostScript figures go on separate pages
% --------------------------------------------------------------------

% Figure 1: Spectrum of intensity fluctuations and single-component fit
\clearpage

\begin{figure}[t]
\epsfxsize=6.0truein
\epsfbox{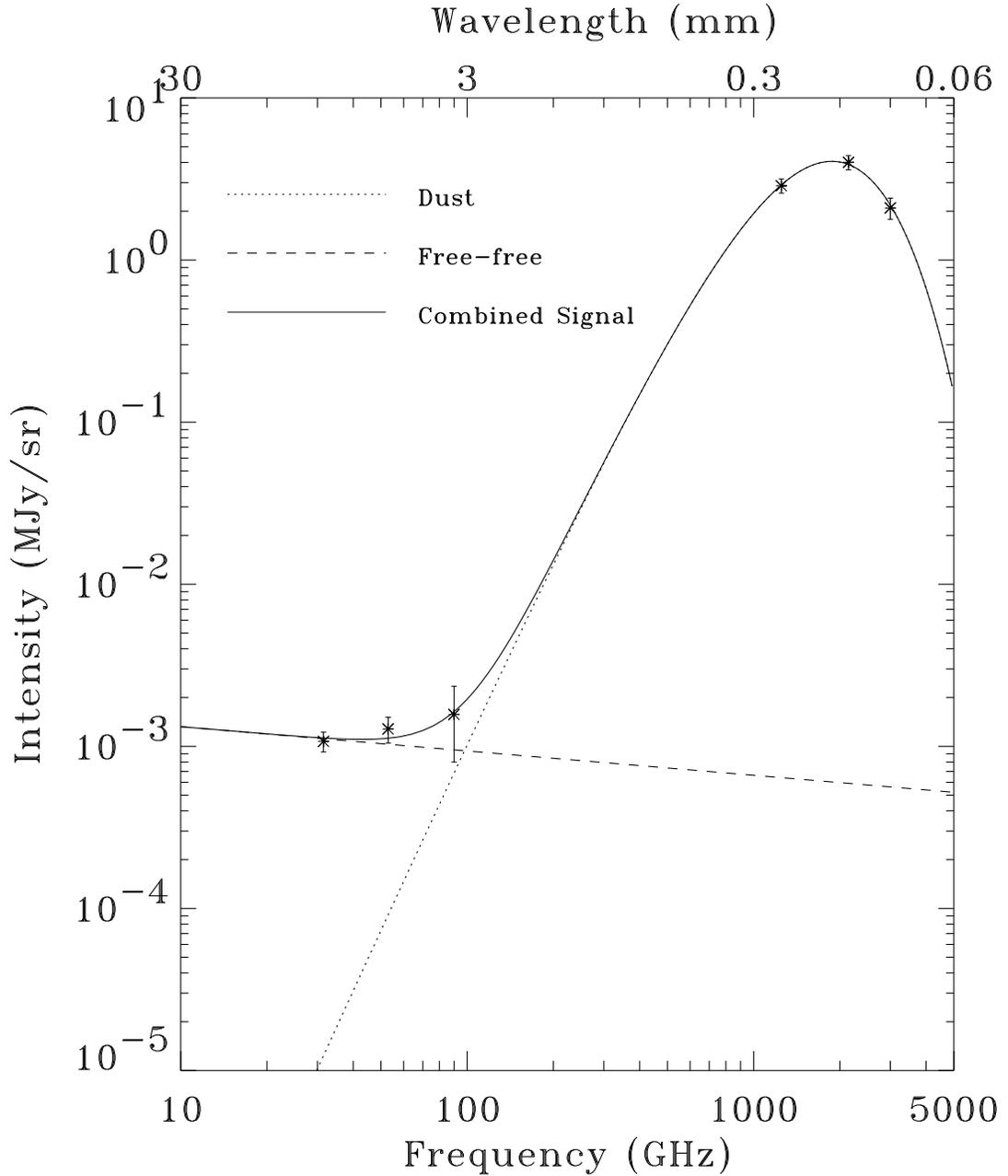}
\caption{
Spectrum of correlated intensity fluctuations 
in the DMR and DIRBE data at $|b| \gt 20\deg$.
Far-infrared points represent the standard deviation of the DIRBE maps
after subtracting a fitted monopole, dipole, and quadrupole,
and include calibration uncertainties.
Long-wavelength points are the values inferred from the cross-correlation of 
the DMR and DIRBE maps (Table 2).
The fitted free-free emission and single-component dust model are also shown.
}
\end{figure}

% Figure 2: Contour plot of dust temperature-emissivity chi-squared
\clearpage

\begin{figure}[t]
\epsfxsize=6.0truein
\epsfbox{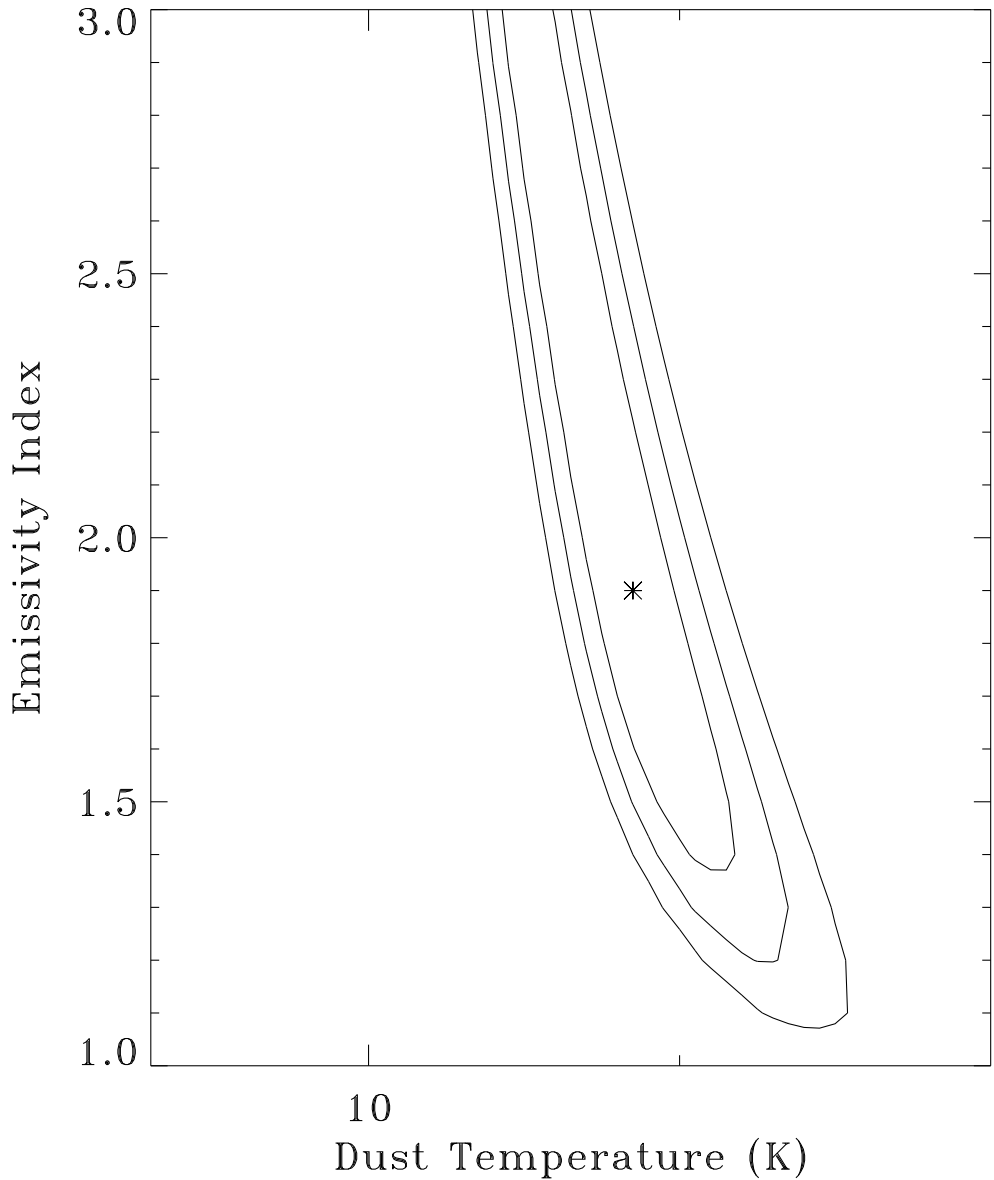}
\caption{
$\chi^2$ contours for the temperature and emissivity index
of a single dust component fitted to the DMR--DIRBE intensity fluctuations
at $|b| \gt 20\deg$.
The asterisk marks the best fitted value.
Free-free emission is fitted separately for each value of the dust emissivity.  
}
\end{figure}

% Figure 3: Frequency spectrum of Galactic emission
\begin{figure}[t]
\epsfxsize=6.0truein
\epsfbox{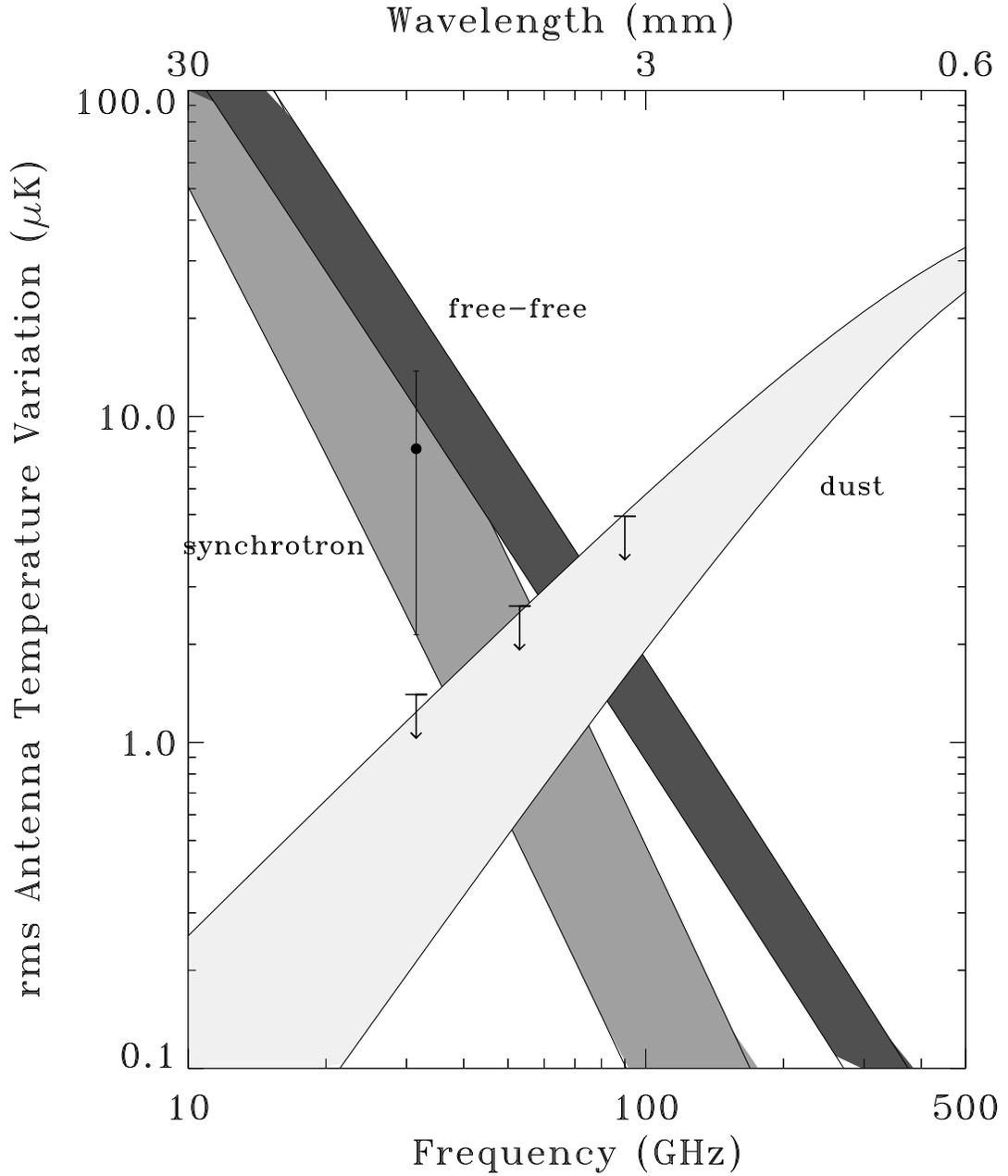}
\caption{
Frequency spectra of high-latitude Galactic emission
({\it rms} fluctuations at $|b| \gt 30\deg$ after quadrupole subtraction
for a 7\deg ~FWHM beam).
Free-free emission dominates synchrotron emission at frequencies
above 20 GHz.
Upper limits on dust emission at millimeter wavelengths
from the DMR--DIRBE cross-correlation,
after correction for free-free emission,
are also shown.
}
\end{figure}

% Figure 4: Power spectrum of free-free emission
\clearpage
\begin{figure}[t]
\epsfxsize=6.0truein
\epsfbox{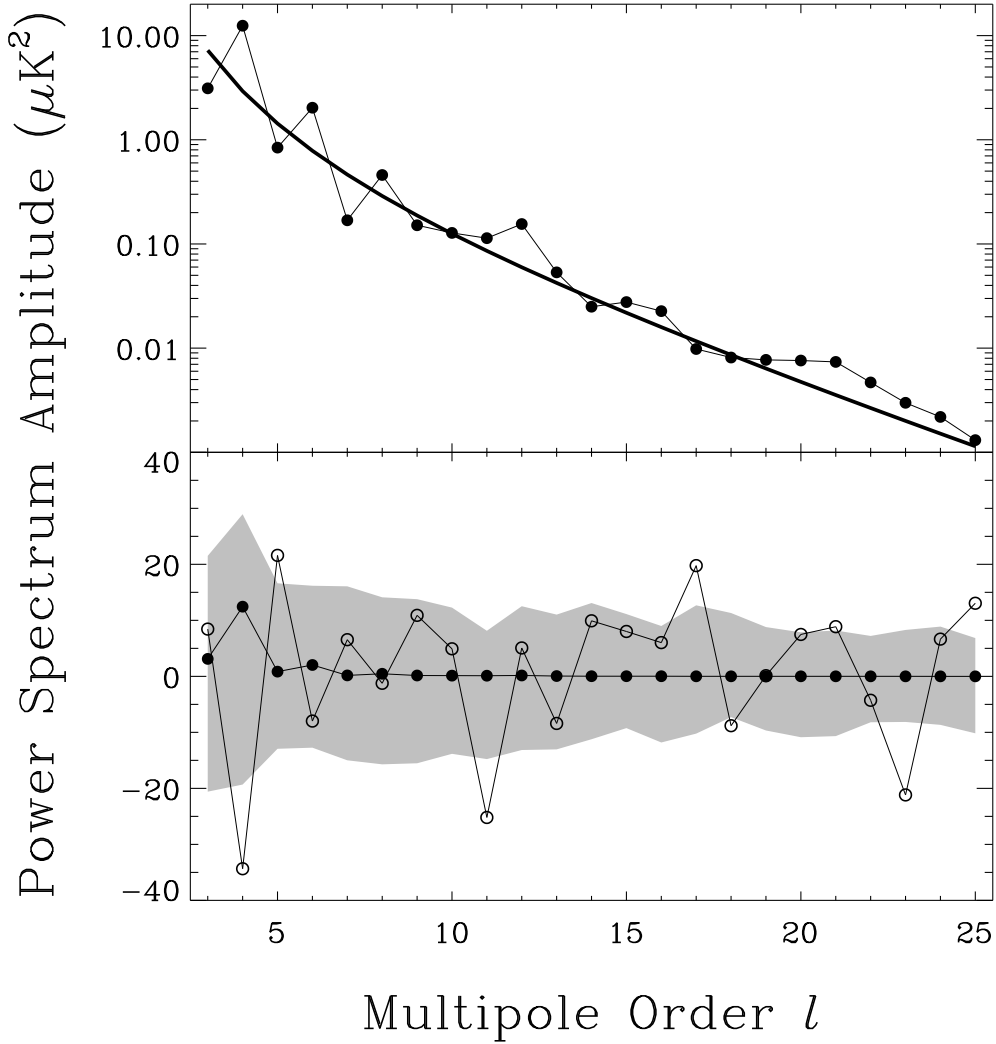}
\caption{
Mean power spectrum of the inferred free-free emission at 53 GHz.
(top) Free-free power spectrum derived from the DMR--DIRBE cross-correlation
at Galactic latitude $|b| \gt 30\deg$.
The filled circles show the DIRBE power spectrum on angular scales
larger than 7\deg, normalized to the correlation coefficient of the
free-free emission at 53 GHz (see text).  
The solid line shows a fitted $\ell^{-3}$ spectrum. 
(bottom) Mean power spectrum of a free-free map
derived from a linear combination of DMR maps
{\it without} specifying the DIRBE maps as the spatial template.
Open circles are the actual power at each multipole order, 
while the grey band shows the 68\% confidence interval
from instrument noise.
Filled circles are the power spectrum of the DMR--DIRBE cross-correlation
from the top panel.
The free-free signal is buried in the instrument noise
unless the spatial template is specified {\it a priori}.
}
\end{figure}

% Th-th-th-that's all, folks!
\end{document}